\documentclass[12pt]{amsart}
\usepackage{amssymb}
\usepackage{euscript}
\usepackage{multicol}
\usepackage{amsmath}
\usepackage{times}

\newtheorem{theorem}{Theorem}[section]

\numberwithin{equation}{section}

\title[Estimates for  eigenvalues]{ Estimates for eigenvalues of the  Schr\"odinger operator with a complex potential}
\author{Oleg Safronov }

\email{osafrono@uncc.edu}


\begin{document}

\maketitle

\begin{abstract}
We study the distribution of  eigenvalues of the Schr\"odinger operator with a complex valued potential $V$. We prove that if $|V|$ decays faster than the Coulomb potential, then all eigenvalues are in a disc of a finite radius. 
\end{abstract}

\section{Introduction}

We consider the Schr\"odinger operator $H = -\Delta + V$ with a complex
potential $V$ and then we study the distribution
of eigenvalues of $H$ in the complex plane.

Our work in this direction  was motivated by
the question of E.B. Davies about an integral estimate for eigenvalues
of $H$ (see \cite{D1} and \cite{AN}). If $d = 1$ then all eigenvalues ¸ of $H$ which do not
belong to ${\Bbb R}_+=[0,\infty)$ satisfy
$$
|\lambda|\leq \frac14\Bigl(\int_{\Bbb R}|V(x)|dx\Bigr)^2.
$$
The question is  whether something  similar    holds in dimension $d\geq2$.
We prove the following result  related directly to this matter.
\begin{theorem}\label{main} 
Let $V:{\Bbb R}^d\mapsto {\Bbb C}$ satisfy the condition
$$
|V(x)|\leq \frac L{(1+|x|^2)^{p/2}}, \qquad 1<p<3,
$$ with a  constant $L>0$. Let $\varkappa=(p-1)/2$ and let $\epsilon>0$ be an arbitrarily  small number that belongs to the intersection of the intervals $(0,(1-\varkappa)/2)\cap(0, 1/2)$.
Then   any  eigenvalue $\lambda\notin {\Bbb R}_+$ of $H$ with $\Re\lambda>0$ satisfies one of the conditions: $${\rm 1) \,\, either} \,\,\,\,\, |\lambda|\leq1$$
or
$$
{\rm 2)} \quad 1 \leq C L\Bigl(|\Re\lambda|^{(\varkappa+2\varepsilon-1)/2}+
|\lambda|^{\varepsilon-1/2}+ \frac {1+|\lambda|^{\varepsilon}}{(|\lambda|-1)}\Bigr)
$$
where the constant $C$ depends on the dimension $d$ and on the parameters $p$ and $\varepsilon$. In particular, it means that  all non-real eigenvalues  are  in a disc of a finite   radius.  
\end{theorem}

The study of eigenvalue estimates  for  operators with a complex potential already has 
a   bibliography.  Besides \cite{D1} and \cite{AN}, we would like to mention
the papers \cite{FLLS} and \cite{LS}.
The main result of \cite{FLLS} tells us,  that for any $t>0$, 
the eigenvalues $z_j$ of $H$
lying outside the   sector $\{z:\ \ |\Im z|<t\ \Re z\}$ satisfy the  estimate 
$$
\sum |z_j|^\gamma\leq C\int |V(x)|^{\gamma+d/2}dx,\qquad \gamma\geq 1,
$$
where  the constant  $C$  depends on $t, \gamma$ and $d$ (see also \cite{LT} for the case  when $V$ is real). 

The paper \cite{LS}  deals with natural question that appears in relation to the main  result of \cite{FLLS}: what estimates are valid for the eigenvalues  situated inside  the conical sector  $\{z:\ |\Im z|<t \Re z\}$, where  the eigenvalues might be close to the positive half-line?
Theorems of the article \cite{LS} provide some  information about the
 rate of accumulation of  eigenvalues to the set ${\Bbb R}_+=[0,\infty)$. 
Namely, \cite{LS} gives  sufficient conditions on 
$V$ that  guarantee convergence of the sum
$$
\sum_{a<\Re z_j<b}|\Im z_j|^\gamma<\infty
$$
for  $0\leq a<b<\infty$.
Moreover, the following result is also proven in \cite{LS}:
\begin{theorem}\label{LS}  Let $V$   be a  function  from
 $L^p({\Bbb R}^d)$, where 
$p\geq d/2$, if $d\geq3$>;  $p>1$, if $d=2$, and $p\geq1$, if $d=1$. 
Then every  eigenvalue $\lambda$ of the operator
$H=-\Delta+V$  with the property $\Re \lambda> 0$ satisfies the estimate
\begin{equation}\label{21t3}
|\Im \lambda|^{p-1}\leq |\lambda|^{d/2-1}C\int_{{\Bbb R}^d} |V|^pdx.
\end{equation}
The constant  $C$ in this inequality depends only on $d$ and $p$.
Moreover, $C=1/2$ for $p=d=1$.
\end{theorem}

\section{Proof of Theorem~\ref{main}}
Consider  first the case $L=1$. 
For the sake of convenience we introduce the notations $W=|V|^{1/2}$ and $l=p/2$. According to the  Birman-Schwinger principle, a number  $\lambda\notin {\mathbb R}_+$ is  an eigenvalue of  the operator  $H=-\Delta+V(x)$  if and only if  the number $-1$ is an eigenvalue of the operator 
$$
X_0=W(-\Delta-\lambda)^{-1}W\frac{V}{|V|}.
$$
Therefore if $\lambda$ is a point of the spectrum of the operator $H$, then $||X_0||\geq1$. On the other side,
since  multiplication by the function $\frac{V}{|V|}$
represents a unitary operator, the   condition $||X_0||\geq1$
implies that the norm of the operator
$$
X=W(-\Delta-\lambda)^{-1}W
$$
is also not less  than 1. 

In order to estimate   the norm of the operator $X$ from above, we consider  its kernel
$$
(2\pi)^{-d}W(x)\int\frac{e^{i\xi(x-y)}}{\xi^2-\lambda}d\xi\,W(y)
$$
It follows from this formula that $X$ can be represented in the form
$$
X=\int_0^\infty\frac{\Gamma_\rho^*\Gamma_\rho}{\rho^2-\lambda}d\rho,
$$
where $\Gamma_\rho$  is the operator mapping $L^2({\mathbb R}^d)$ into $L^2({\mathbb S}_\rho)$, and  ${\mathbb S}_\rho$ is the sphere of radius $\rho$ with the center at the point $0$:
$$
\Gamma_\rho
u(\theta)=(2\pi)^{-d/2}\int_{{\mathbb R}^d}e^{-i\rho(\theta x)}W(x)u(x)dx
$$
The main properties  of this operator follow   from Sobolev's 
embedding theorems. Suppose that $W(x)\leq (1+|x|^2)^{-l/2}$ and $u\in L^2({\mathbb R}^d)$. Then the Fourier   transformation  of the   function $W(x)u(x)$ belongs to the class $H^l({\mathbb R}^d)$, moreover  the norm $||\hat{Wu}||_{H^l}$ is estimated by the norm $||u||_{L^2}$.
According to Sobolev's  theorems, the embedding of the class $H^l({\mathbb R}^d)$ into 
the class   $L^2({\mathbb S}_\rho)$ is continuous under the condition $l>1/2$.
Moreover, the norm of the embedding  operator depends in a weak manner on the parameter $\rho\geq1$.
Indeed, suppose that the inequality 
$$
\int_{{\mathbb S}_1}|\phi(\theta)|^2
d\theta\leq C\int_{{\mathbb R}^d}\Bigl(|\nabla^l\phi|^2+|\phi|^2\Bigr)dx
$$
holds for any function 
$\phi\in H^l({\mathbb R}^d)$. 
Then setting $\phi(x)=u(\rho x)$ we obtain that
$$
\int_{{\mathbb S}_1}|u(\rho\theta)|^2
d\theta\leq C\int_{{\mathbb R}^d}\Bigl(\rho^{2l}|\nabla^l u(\rho x)|^2+|u(\rho x)|^2\Bigr)dx.
$$
Multiplying both sides of this inequality by $\rho^{d-1}$, we  obtain that
$$
\int_{{\mathbb S}_\rho}|u(x)|^2
dS\leq C\int_{{\mathbb R}^d}\Bigl(\rho^{2l-1}|\nabla^l u( x)|^2+\rho^{-1}|u( x)|^2\Bigr)dx.
$$

If $l$ is close to $1/2$  then $\rho^{2l-1}$ practically behaves as a constant.
Anyway, without loss of generality we can assume that for $\rho>1$
$$\int_{{\mathbb S}_\rho}|u(x)|^2
dS\leq C_\varepsilon\rho^{2\varepsilon}||u||^2_{H^l} 
$$ where $\varepsilon>0$ is an arbitrary small number.  It implies that
\begin{equation}\label{1}
||\Gamma_\rho||\leq C_\varepsilon \rho^\varepsilon,\qquad \rho\geq 1.
\end{equation}
 Moreover,
 $\Gamma_\rho$ depends continuously on   the parameter $\rho$ in the following sense.
 Let us introduce  the operator $U_\rho$  that transforms  functions on the sphere ${\mathbb S}_\rho$ into functions on the sphere ${\mathbb S}={\mathbb S}_1$. 
 according to the rule
 $$
 U_\rho u(\theta)=u(\rho \theta) \rho^{(d-1)/2}.
 $$
 This operator is unitary and therefore its norm equals 1.
 Define now the operator $Y_\rho=U_\rho \Gamma_\rho$.
 Our   statement is  that
 $$
 ||Y_{\rho'}-Y_\rho||\leq C |\rho'-\rho|^{\alpha}\rho^\delta(\rho^\varepsilon+(\rho')^\varepsilon)
 $$
 where $\alpha<l-1/2$, $\ \delta=l-\alpha-1/2$ and $\rho'>\rho\geq1$.
 Our arguments   are similar to  those we  used  in the proof of  the inequality \eqref{1}. If we assume that   the inequality
$$
\int_{{\mathbb S}_1}|\phi((1+h)\theta)-\phi(\theta)|^2
d\theta\leq Ch^{2\alpha}\int_{{\mathbb R}^d}\Bigl(|\nabla^l\phi|^2+|\phi|^2\Bigr)dx
$$
holds for any function 
$\phi\in H^l({\mathbb R}^d)$. 
Then the substitution  $\phi(x)=u(\rho x)$ will lead to the inequality 
$$
\int_{{\mathbb S}_1}|u((1+h)\rho\theta)-u(\rho\theta)|^2
d\theta\leq Ch^{2\alpha}\int_{{\mathbb R}^d}\Bigl(\rho^{2l}|\nabla^l u(\rho x)|^2+|u(\rho x)|^2\Bigr)dx.
$$
Multiplying both sides of this inequality by $\rho^{d-1}$ and denoting $\rho'=(1+h)\rho$, we  obtain that
$$
\int_{{\mathbb S}_\rho}|u(\rho^{-1}\rho'x)-u(x)|^2
dS\leq C|\rho'-\rho|^{2\alpha}\int_{{\mathbb R}^d}\Bigl(\rho^{2\delta}|\nabla^l u( x)|^2+\rho^{-2l}|u( x)|^2\Bigr)dx.
$$
provided that $\rho'>\rho\geq1$.  This  leads to
$$
||\Bigl(\frac{\rho}{\rho'}\Bigr)^{(d-1)/2}Y_{\rho'}-Y_{\rho}||\leq C|\rho'-\rho|^{\alpha}\rho^{\delta}.
$$
We apply  now the triangle inequality to estimate  the norm of the difference 
$Y_{\rho'}-Y_{\rho}$ for $\rho'>\rho\geq 1$
$$
||Y_{\rho'}-Y_{\rho}||\leq \Bigl|\Bigl(\frac{\rho}{\rho'}\Bigr)^{(d-1)/2}-1\Bigr|\,
||Y_{\rho'}||+C|\rho'-\rho|^{\alpha}\rho^{\delta}\leq C_0 (\rho^\varepsilon+ (\rho')^\varepsilon)|\rho'-\rho|^{\alpha}\rho^{\delta}.
$$
To be more convincing, we  mention that 
$$
\Bigl|\Bigl(\frac{\rho}{\rho'}\Bigr)^{(d-1)/2}-1\Bigr|\leq \min\{2^{-1}(d-1)|\rho'-\rho|,\, 2\}.
$$
Introduce now  the notation $G_\rho=\Gamma^*_\rho \Gamma_\rho$.
Obviously, $G_\rho$  aslo has  representation
$G_\rho =Y^*_\rho Y_\rho $.
Consequently, 
$$
||G_{\rho'}-G_{\rho}||\leq ||Y^*_{\rho'}-Y^*_{\rho}||\cdot||Y_{\rho'}||+||Y^*_{\rho}||\cdot||Y_{\rho'}-Y_{\rho}||\leq C( \rho^\varepsilon+ (\rho')^\varepsilon)^2|\rho'-\rho|^{\alpha}\rho^{\delta}.
$$
Let us summarize the results. The operator $X$ can be written in the form
$$
X=\int_0^\infty \frac{G_\rho d\rho}{\rho^2-\lambda},
$$
where 
$$
||G_\rho||\leq C\rho^{2\varepsilon},\qquad \rho\geq1,
$$
and
$$
||G_{\rho'}-G_{\rho}||\leq C ( \rho^\varepsilon+ (\rho')^\varepsilon)^2|\rho'-\rho|^{\alpha}\rho^{\delta},\qquad \rho'>\rho\geq1.
$$
Now, since  the integral representation for the operator $X$  can be also   rewritten in the form
$$
X=\int_1^\infty \frac{(G_\rho-G_\tau )d\rho}{\rho^2-\lambda}+\int_1^\infty \frac{G_\tau d\rho}{\rho^2-\lambda}+ W(-\Delta-\lambda)^{-1}E[0,1]W
$$
where $\tau=|\Re\lambda|^{1/2}$ and $E[0,1]$ is the spectral projection of 
the operator $-\Delta$ corresponding to the interval $[0,1]$, we obtain that
$$
||X||\leq \int_1^\infty \frac{||G_\rho-G_\tau || d\rho}{|\rho^2-\lambda|}+\frac{\pi||G_\tau ||}{2|\lambda|^{1/2}}+ \frac{||V||_{L^\infty}+||G_\tau||}{(|\lambda|-1)},
$$ for $ |\lambda|>1.$
Consequently, 
$$
||X||\leq  
C\int_0^\infty 
\frac {|\rho-\tau|^\alpha ( \rho^\delta+|\Re\lambda|^{\delta/2})( \rho^\varepsilon+|\Re\lambda|^{\varepsilon/2})^2}
{|\rho^2-\Re\lambda|}d\rho+
C \frac {\tau^{2\varepsilon }}{|\lambda|^{1/2}}+ \frac {||V||_{L^\infty}+C\tau^{2\varepsilon}}{(|\lambda|-1)},
$$
which leads to
$$
1\leq ||X||\leq  C \Bigl(|\Re\lambda|^{(\alpha+\delta+2\varepsilon-1)/2}+
|\lambda|^{\varepsilon-1/2}+ \frac {1+|\lambda|^{\varepsilon}}{(|\lambda|-1)}\Bigr).$$
We proved the statement of the theorem for the case $L=1$. If $L\neq1$ then this inequality takes the form
$$
1 \leq C L\Bigl(|\Re\lambda|^{(\alpha+\delta+2\varepsilon-1)/2}+
|\lambda|^{\varepsilon-1/2}+ \frac {1+|\lambda|^{\varepsilon}}{(|\lambda|-1)}\Bigr).$$
The proof is completed. $\,\,\,\,\,\Box$

\end{document}